\documentclass[%
 aip,
 amsmath,amssymb,
 reprint,%
]{revtex4-1}

\usepackage{graphicx}
\usepackage{dcolumn}
\usepackage{bm}

\usepackage[utf8]{inputenc}
\usepackage[T1]{fontenc}
\usepackage{mathptmx}
\usepackage{etoolbox}
\usepackage{color}

\makeatletter
\def\@email#1#2{%
 \endgroup
 \patchcmd{\titleblock@produce}
  {\frontmatter@RRAPformat}
  {\frontmatter@RRAPformat{\produce@RRAP{*#1\href{mailto:#2}{#2}}}\frontmatter@RRAPformat}
  {}{}
}%
\makeatother
\begin{document}

\preprint{AIP/123-QED}

\title[Mechanism of generating collisionless shock in magnetized gas plasma]{Mechanism of generating collisionless shock in magnetized gas plasma driven by laser-ablated target plasma}

\author{S. Matsukiyo}
\affiliation{Faculty of Engineering Sciences, Kyushu University \\
6-1 Kasuga-Koen, Kasuga, Fukuoka 816-8580, Japan}
\affiliation{Interdisciplinary Graduate School of Engineering Sciences, Kyushu University \\
6-1 Kasuga-Koen, Kasuga, Fukuoka 816-8580, Japan}
\affiliation{Institute of Laser Engineering, Osaka University \\
2-6, Yamadaoka, Suita, Osaka 565-0871, Japan}
\email{matsukiy@esst.kyushu-u.ac.jp}

\author{K. Oshida}
\affiliation{Interdisciplinary Graduate School of Engineering Sciences, Kyushu University \\
6-1 Kasuga-Koen, Kasuga, Fukuoka 816-8580, Japan}

\author{S. Isayama}
\affiliation{Faculty of Engineering Sciences, Kyushu University \\
6-1 Kasuga-Koen, Kasuga, Fukuoka 816-8580, Japan}

\author{R. Yamazaki}
\affiliation{Department of Physical Sciences, Aoyama Gakuin University \\
5-10-1 Fuchinobe, Sagamihara, Kanagawa 252-5258, Japan}
\affiliation{Institute of Laser Engineering, Osaka University \\
2-6, Yamadaoka, Suita, Osaka 565-0871, Japan}

\author{T. Morita}
\affiliation{Faculty of Engineering Sciences, Kyushu University \\
6-1 Kasuga-Koen, Kasuga, Fukuoka 816-8580, Japan}

\author{T. Takezaki}
\affiliation{Faculty of Engineering, University of Toyama\\
3190, Gofuku, Toyama 930-8555, Japan}

\author{Y. Kuramitsu}
\affiliation{Graduate School of Engineering, Osaka University\\
2-1, Yamadaoka, Suita, Osaka 565-0871, Japan}

\author{S. J. Tanaka}
\affiliation{Department of Physical Sciences, Aoyama Gakuin University \\
5-10-1 Fuchinobe, Sagamihara, Kanagawa 252-5258, Japan}

\author{T. Sano}
\affiliation{Institute of Laser Engineering, Osaka University \\
2-6, Yamadaoka, Suita, Osaka 565-0871, Japan}

\author{K. Tomita}
\affiliation{Division of Quantum Science and Engineering, Hokkaido University\\
Sapporo 060-8628, Japan}

\author{Y. Sakawa}
\affiliation{Institute of Laser Engineering, Osaka University \\
2-6, Yamadaoka, Suita, Osaka 565-0871, Japan}

%
%
\date{\today}

\begin{abstract}
Mechanism of generating collisionless shock in magnetized gas plasma driven 
by laser-ablated target plasma is investigated by using one-dimensional full 
particle-in-cell simulation. The effect of finite injection time of target 
plasma, mimicking finite width of laser pulse, is taken into account. It was 
found that the formation of a seed-shock requires a precursor. The 
precursor is driven by gyrating ions, and its origin varies depending on 
the injection time of the target plasma. When the injection time is short, 
the target plasma entering the gas plasma creates a precursor, otherwise, 
gas ions reflected by the strong piston effect of the 
target plasma create a precursor. The precursor compresses the background 
gas plasma, and subsequently, a compressed seed-shock forms in the gas 
plasma. The parameter dependence on the formation process and propagation 
characteristics of the seed-shock was discussed. It was confirmed that 
the seed-shock propagates through the gas plasma exhibiting behavior 
similar to the shock front of supercritical shocks.
\end{abstract}

\maketitle


\section{\label{sec:intro}Introduction}

While collisionless shocks have long been studied as powerful accelerators 
of cosmic rays, their acceleration mechanisms remain highly complex 
and largely veiled. Studies of collisionless shocks have traditionally 
employed various methods such as in-situ and remote observations in 
space, theoretical modeling, and numerical simulations. Recently, 
in addition to these methods, efforts to generate magnetized 
collisionless shocks using high-power laser facilities in laboratory 
experiments have garnered significant attention \cite{matsukiyo22,yamazaki22,
yao21,schaeffer19,schaeffer17,kuramitsu16,schaeffer15,schaeffer14,
niemann14}. 

Among a number of methods of shock generation, Yamazaki et al. (2022)\cite{yamazaki22} 
and Matsukiyo et al. (2022)\cite{matsukiyo22} 
proposed a method to generate a shock in a homogeneously magnetized ambient 
gas plasma with uniform Alfv\'{e}n velocity. In the experiment, a 
solid aluminum target irradiated with a laser is ablated and a 
surrounding nitrogen gas filling uniformly an entire chamber is ionized 
by the strong radiation emitted by the laser-target interaction. The 
ionized gas plasma initially at rest is uniformly magnetized in sufficiently 
large volume for a long time by using a Helmholtz-like coil. 
Although they observed developing shock (seed-shock hereafter), propagating in the 
gas plasma, detailed formation process of the seed-shock in this system has 
not been fully understood. 

The complexity arises from the highly intricate interaction between 
gas plasma and target plasma. Since it is nearly impossible to separately 
measure the two plasmas in experiments, comparison with numerical 
simulations becomes crucial. However, there has not been a well-established 
computational method that accurately replicates these early-stage shock 
formations. One commonly used numerical simulation in this field is
the Particle-in-Cell (PIC) simulation \citep{birdsall04}. 
A number of methods of shock 
generation using the PIC simulation have been proposed so far 
\textcolor{black}{
(e.g. \onlinecite{pongkitiwanichakul24,zhang21,moreno20,schaeffer20,dieckmann18} and reviewed in Ref.~\onlinecite{burgess15}). 
}
Normally, as enough time 
elapses, the influence of initial conditions and boundary 
conditions can be ignored, so differences in shock generation 
methods are not a concern. However, to accurately simulate 
the initial stages of shock generation observed in 
laser experiments, considerable attention needs to be paid to 
the initial setup of the calculations. 

Fox et al. (2013)\cite{fox18} introduced a localized heating 
operator to refine and incorporate the effects of laser ablation. 
This involves setting up a high-density region mimicking the target 
at one end of the system (with a constant density over time) and 
imparting a sufficiently high plasma temperature only to that specific 
area. This allows the target plasma to thermally expand into its 
surroundings. In their approach, it is necessary to handle a very 
high-density plasma in a narrow region, which results in high 
computational costs. Matsukiyo et al. (2022)\cite{matsukiyo22} employed 
a different method to reproduce similar effects in one-dimensional 
simulations. They injected a high-temperature, high-density target 
plasma into a system filled with uniform background gas plasma at a 
specific point for a finite duration. This eliminates the need to 
deal with target material that does not interact with the background 
gas plasma, thereby reducing computational costs.
In the 
simulation they showed that injected target ions gyrate around ambient 
magnetic field in a gas plasma and turned back after a quarter of their 
gyro period. During this, a background gas ions are dragged 
(or accelerated) by the target ions so that the gas ions are gradually 
accumulated and compressed to form a shock-like steepened density 
profile afterwards. They called this steepened structure a developing 
shock. However, it is unclear how universal the process is. If the 
mechanism for generating a shock in experiments is elucidated, it would 
enable us to devise means to generate a shock in a short period 
of time, making it possible to measure a more developed shock.

In this paper we use the same simulation method as 
Ref.~\onlinecite{matsukiyo22} to investigate the mechanism of shock formation 
in a magnetized gas plasma interacting with an ablated target plasma. We 
will show that the mechanism of shock formation changes depending 
on the injection time of target plasma. Moreover, it will be 
demonstrated that the time it takes to form a sufficiently steep 
density structure depends on the strength of the ambient magnetic 
field.
To distinguish the structure we mainly discuss in this paper with 
usually discussed a well developed shock, we call the early stage 
steepened density or magnetic field structure discussed here a 
seed-shock, which does not accompany well developed steady and 
long enough downstream state. A seed-shock is equivalent to the 
structure named a developing shock in Ref.~\onlinecite{matsukiyo22}.

The paper is organized as follows. In section \ref{sec:setting} 
simulation settings are explained. The mechanism of seed-shock formation 
is discussed in section \ref{sec:mechanism}. The parameter 
dependence of seed-shock properties is shown in section \ref{sec:dependence}. 
Then, discussions are given in section \ref{sec:discussion}.

\section{\label{sec:setting}Simulation settings}

\textcolor{black}{
We perform one-dimensional full particle-in-cell simulations using the 
same simulation code as in Ref.~\onlinecite{matsukiyo22}.
}
Initially a thermal gas plasma consisting of electrons and ions 
is uniformly distributed in the system ($-0.3 L_x \le x \le 0.7 L_x$). 
The valence of gas ions is assumed to be $Z_G=1$ throughout this study. 
The ambient magnetic field is applied in $z-$direction. The velocity 
distribution function is Maxwellian for both electrons and ions. 
In most of the following runs, $L_x \approx \textcolor{black}{130} c/\omega_{Gi}$, where $c$ 
is the speed of light and $\omega_{Gi}$ denotes ion plasma frequency. 
\textcolor{black}{
The number of spatial grids is $9 \times 10^4$, and the number 
of particles per cell is 256 for each species.
}
We postulate that a target plasma is created for a finite period of 
time with a constant rate. If a solid target is thick enough, 
the created target electrons may have a half-Maxwellian 
distribution function in which there are no electrons having negative 
velocity in $v_x (<0)$. The target ions, on the other hand, may have 
a bulk velocity, $v_{in}$, to compensate the current produced by 
the half-Maxwellian electrons. With this assumption, the temperature 
of the target electrons can be written as $T_e = 2m_e v^2_{in}$. 
We further assume that the temperature of target ions is the same 
as the electrons, $T_i=T_e$, and ion distribution function is 
shifted-Maxwellian. The target plasma is injected at $x=0$ during 
$0 \le t \le T_{in}$. 
\textcolor{black}{
It carries the magnetic field generated through the laser-target 
interaction, which dynamically changes based on the motion of the 
target plasma. 
}
In experiment, this magnetic field, 
known as the Biermann battery field \citep{stamper71,stamper78,yates82}, is generated in a ring-like shape 
around the laser spot on the target surface and expands over time 
along with the target plasma. In this one-dimensional simulation 
we assume that the direction of this magnetic field is in 
$z-$direction. One of the objectives of this paper is to elucidate 
the effects of finite injection time of target plasma in the generation 
of a shock in a gas plasma. This is equivalent to accounting for the 
effects of finite laser irradiation time in experiments (Note that 
total laser power is proportional to $T_{in}$.).

In Runs 1-6, for the gas plasma, 
the ion-to-electron mass ratio is $m_G/m_e=100$, the ratio of 
electron cyclotron frequency to plasma frequency 
$\Omega_{Ge}/\omega_{Ge}=0.05$, the electron and ion betas 
$\beta_{Ge}=\beta_{Gi}=0.1$, respectively. For the target plasma, 
the valence of ions is $Z_T=6$, and the ion-to-electron mass ratio 
$m_T/m_e=200$. The density ratio of target electrons to gas electrons 
is $N_{Te}/N_{Ge}=7$ and the magnetic field carried by the target 
plasma is 6 times the ambient field ($\Omega_{Te}/\Omega_{Ge}=6$). 
The mass ratio of target ions to gas ions is $m_T/m_G=2$ which is 
close to that of aluminum (target) to nitrogen (gas) ($\approx 1.93$). 
In the subsequent sections, calculated values are presented in both 
normalized and unit-attached forms. Normalization is based on the 
scale of gas ions, i.e., time is to inverse ion gyro frequency, 
$\Omega^{-1}_{Gi}$, velocity is to Alfv\'{e}n velocity, $v_A$, 
and space is to ion inertial length, $v_A/\Omega_{Gi}$, respectively. 
When converting to values with units attached, typical values 
from the Gekko XII experiment are used as normalization constants 
for these ions. Since we have used unrealistic 
ion-to-electron mass ratio, electron dynamics such as gyro motion 
and Debye shielding have relatively larger scale lengths. However, 
they are still much smaller than ions' so that electron scale 
phenomena in the simulation here may not affect significantly in 
ion dynamics. We will confirm this later. 
\textcolor{black}{
It should also be noted that in Ref.~\onlinecite{matsukiyo22}, 
the same simulation code as well as the similar parameters 
used here were employed, and the results 
were found to be in good agreement with the experimental data 
discussed in it.
}

The injection speed of target ions using the Gekko XII experiment 
at the Institute of Laser Engineering in Osaka University is typically 
a little less than 1,000 km/s for the laser intensity of 
$\sim 10^{13} \rm{W/cm^2}$. Here, we assume that the injection 
speed is 850 km/s. In Runs 1-3, we assume that $v_{in}=21.7 v_A$, 
where $v_A$ denotes the Alfv\'{e}n velocity in the gas plasma. This 
corresponds to the case of the ambient magnetic field strength 
being 4.4 T for $1.4 \times 10^{18} \rm{cm}^{-3}$ nitrogen ions, $N^+$. 
In Runs 4-6, on the other hand, $v_{in}=15.9 v_A$ is 
assumed, corresponding to the case with 6 T ambient magnetic field 
and the same ion density. 
For the both cases, five different values of $T_{in}$ are used. 
The above values are summarized in \textcolor{black}{Table} \ref{tab:my_label}.

\begin{table}
    \centering

    \begin{tabular}{|c|c|c|c|c|c|}
    \hline
    $B_0 \backslash T_{in}$ & 0.65 [ns] & 1.3 [ns] & 2.6 [ns] & 3.9 [ns] & 5.2 [ns]\\
    \hline
        4.4 [T] & Run 1 & Run 2 & Run 3 & $\circ$ & $\circ$ \\
        $(\Omega_{Gi}T_{in})$ & (1.92E-2) & (3.92E-2) & (7.83E-2) & (1.17E-1) & (1.57E-1) \\
        $(\Omega_{Ti}T_{in})$ & (5.87E-2) & (1.17E-1) & (2.35E-1) & (3.52E-1) & (4.70E-1) \\
   \hline
        6.0 [T] & Run 4 & Run 5 & Run 6 & $\circ$ & $\circ$\\
        $(\Omega_{Gi}T_{in})$ & (2.67E-2) & (5.34E-2) & (1.07E-1) & (1.60E-1) & (2.14E-1) \\
        $(\Omega_{Ti}T_{in})$ & (8.01E-2) & (1.60E-1) & (3.20E-1) & (4.81E-1) & (6.41E-1) \\
   \hline
    \end{tabular}
    \caption{Values of ambient magnetic field strength, $B_0$, and 
    injection time, $T_{in}$, examined.}
    \label{tab:my_label}
\end{table}

\section{\label{sec:mechanism}Mechanism of shock formation}
%

\subsection{Spatio-temporal evolution of density}

\begin{figure}[ht]
\includegraphics[clip, width=1.0\columnwidth]{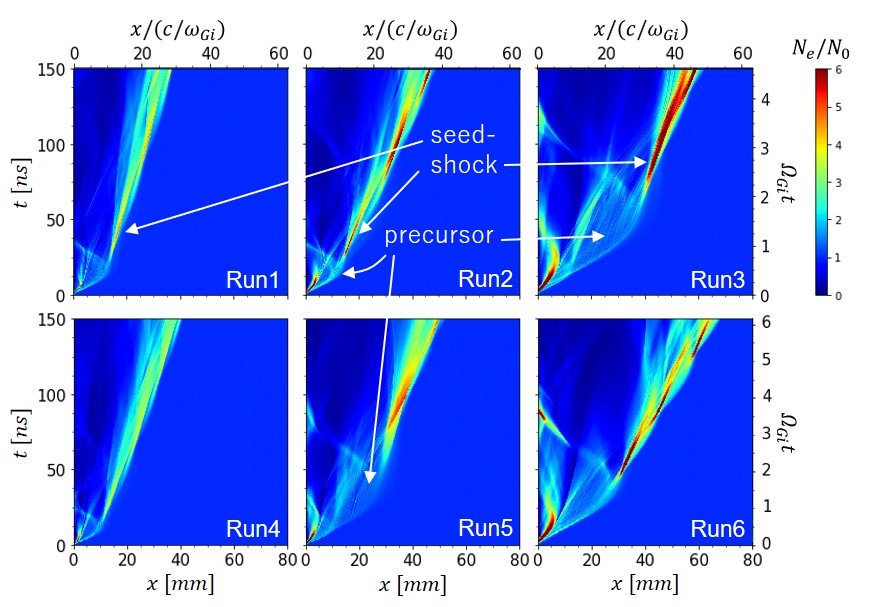}
\caption{\label{fig:spacetime}Spatio-temporal evolution of electron density 
in Runs 1-6.}
\end{figure}
%
Fig.\ref{fig:spacetime} shows the spatio-temporal evolution of 
electron density for Runs 1-6. In each case the red colored high 
density structure initially occurs near $x=0$ and propagates while 
forming an arc in the $x-t$ space. In front of it, a complex light 
blue or yellow-green structure, hereafter we call it a precursor, 
is formed. When it turns around, another red (or yellow for Run 4) 
colored compressed structure appears and propagates forward. 
\textcolor{black}{
Here, we define a precursor as the structure arching out into upstream, 
observed just before the compressed structure propagating forward 
begins to grow.
}

As will be discussed later, the compressed structure propagating 
forward grows to become a shock in each case. 
\textcolor{black}{
Therefore, we will refer to this structure as a seed-shock.
}
The approximate time 
when \textcolor{black}{a seed-shock} first becomes visible for $B_0=4.4$ T is around 
30 ns in Runs 1 and 2, while it is around 60 ns in Run 3. 
Similarly, the corresponding time for $B_0=6$ T is around 20 ns in 
Run 4, while it is around 50 ns or less in Runs 5 and 6. There 
is a big jump between Run 2 and Run 3 (Run 4 and Run 5), implying that 
the mechanism of shock formation changes here. 
\textcolor{black}{
From Table I, the change may occur at $\Omega_{Gi}T_{in} \sim 4-5 \times 10^{-2}$ 
($\Omega_{Ti}T_{in} \sim 0.1$, where $\Omega_{Ti}$ is target ion gyro frequency).
}

\subsection{Seed-shock formation: Run 2}

\begin{figure}[ht]
\includegraphics[clip, width=1.0\columnwidth]{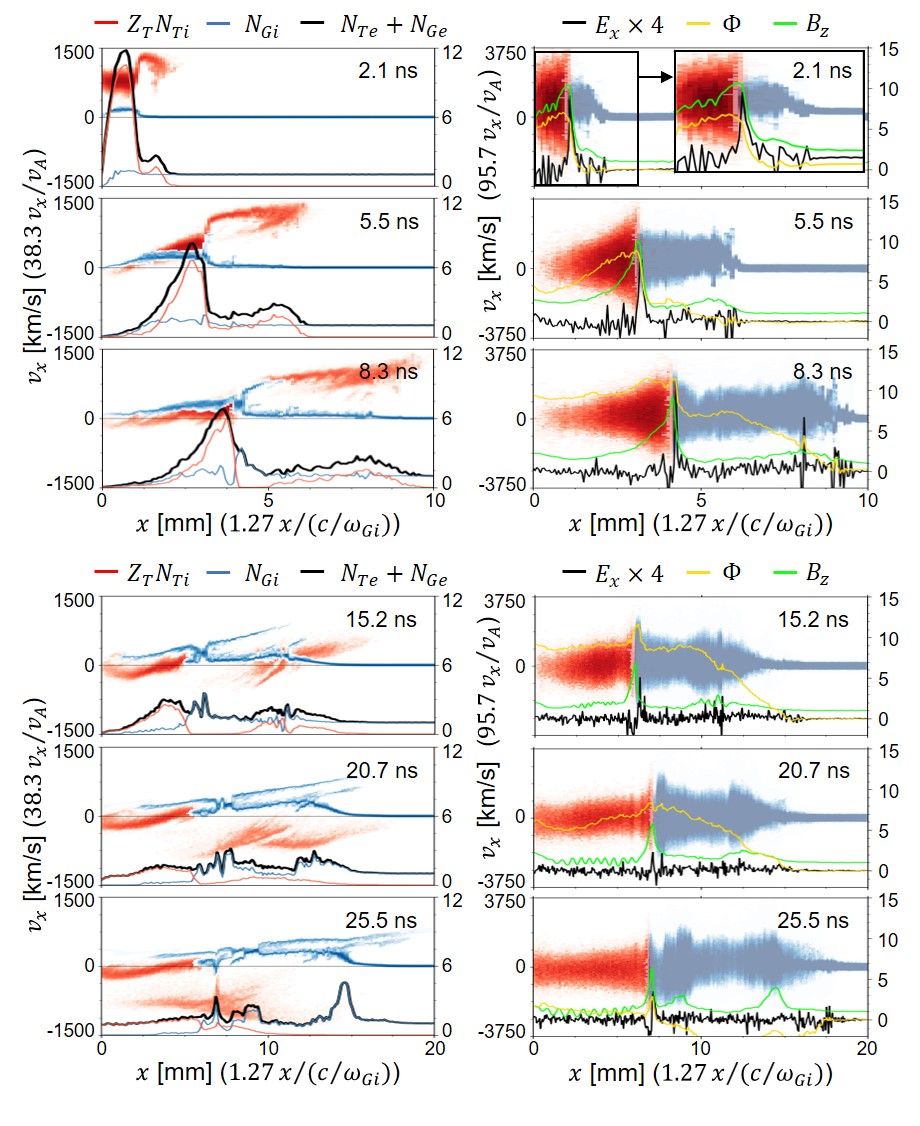}
\caption{\label{fig:run2_1st}The phase space distribution of target and gas 
plasmas, along with field profiles at various time points for Run 2. In each 
time, the left panel shows ion phase space (reddish color: target ions, 
bluish color: gas ions) and charge density profiles (red line: target ions, blue 
line: gas ions, black line: total electrons), while the right panel represents 
electron phase space (reddish color: target electrons, the bluish 
color: gas electrons) and field profiles (green line: $B_z$, black 
line: $E_x$, yellow line: $\Phi$).}
\end{figure}
\begin{figure}[ht]
\includegraphics[clip, width=1.0\columnwidth]{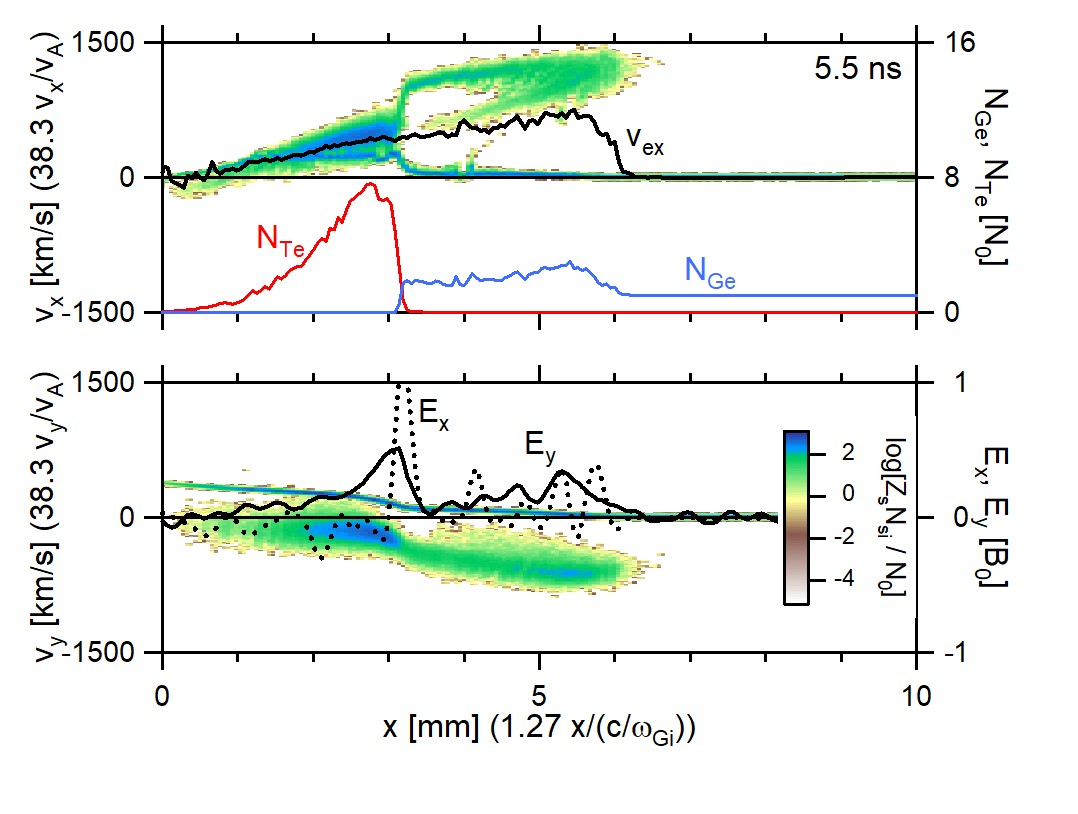}
\caption{\label{fig:vex} Top: $v_x - x$ phase space distribution 
of \textcolor{black}{charge density of} target and gas ions at $t=5.5$ ns 
and profiles of electron bulk 
velocity ($v_{ex}$: black line), density of target ($N_{Te}$: red line) 
and gas ($N_{Ge}$: blue line) 
electrons. The densities are normalized to upstream gas density, $N_0$. 
Bottom: $v_y - x$ phase space distribution 
of target and gas ions at $t=5.5$ ns and profiles of electric 
field components of $E_x$ (dotted line) and $E_y$ (solid line). 
The fields are normalized to ambient magnetic field, $B_0$.}
\end{figure}

%
Fig.\ref{fig:run2_1st} denotes the evolution of phase space and field profiles in 
the early stage of Run 2. In each time, the left panel shows the phase space 
density of target (gas) ions in reddish (bluish) color scale, and the ion 
density profiles of target and gas ions are denoted by red and blue lines, 
where the target ion density is multiplied by the valence of ions. The 
total electron density is indicated by black solid line. In the right panel, 
the phase space density of target (gas) electrons are denoted by reddish 
(bluish) color scale, while the green, black, and yellow lines indicate 
the profiles of $B_z$, $E_x$, and potential $\Phi$, respectively. 
As time passes, the gas plasma is pushed by the target plasma so that a 
majority of injected target ions are decelerated and the gas ions are 
accelerated ($t=2.1-5.5$ ns). Therefore, more ions are accumulated 
near the interface. 
Fig.\ref{fig:vex} provides a more detailed view of this situation at $t=5.5$ ns. 
The upper panel shows \textcolor{black}{the $v_x-x$ phase space distribution 
of ions' charge density} in color 
scale, the bulk velocity of the electrons in the $x-$direction ($v_{ex}$) 
with a black line, and the densities of the target electrons ($N_{Te}$) 
and gas electrons ($N_{Ge}$) with red and blue lines, respectively. The 
target and gas electrons are spatially well separated, as they are tied with 
magnetic field lines of each origin. In the lower panel, the ion's $v_y-x$ 
phase space distribution is shown in color scale, and the $E_x$ and $E_y$ 
components of the electric field are represented by a black dotted line 
and a solid black line, respectively. The magnetized incident target 
plasma carries a convective electric field ($E_y$), which accelerates 
the gas plasma in the $y-$direction while the magnetic field causes it 
to gain velocity in the $x-$direction (the ${\bf E} \times {\bf B}$ motion). 
The motion of gas ions corresponding to this process can be observed in 
the region $x<3$ mm in the figure. At the same time, the target ions 
decelerate due to the reaction force. As a result, ions gradually 
accumulate in this boundary region, forming a potential maximum 
(Fig.\ref{fig:run2_1st}). To the left of this maximum ($x<3$ mm), the 
spatial gradient of the potential becomes positive, leading to a negative 
$E_x$, which also contributes to the deceleration of the incoming target 
ions. Conversely, to the right of the potential maximum ($x>3$ mm), 
a positive $E_x$ is generated due to the negative spatial gradient of 
the potential. Some of the target ions that enter this region are 
accelerated by this $E_x$. In response to the $x-$direction current 
generated by the accelerated target ions, the gas electrons are 
accelerated in the $x-$direction (refer to the black line ($v_{ex}$) 
in the upper panel of Fig.\ref{fig:vex}). The bulk velocity of these 
electrons induces an electric field in the $y-$direction through Ohm's 
law, with $-({\bf v}_e \times {\bf B})/c$ (see the lower panel of 
Fig.\ref{fig:vex} for $x>3$ mm). This accelerates the gas ions in 
the $y-$direction, causing them to gain velocity in the $x-$direction 
through the magnetic field (${\bf E} \times {\bf B}$ motion). 
The above describes the initial interaction between the target plasma 
and the gas plasma. If we define the region occupied by the gas 
electrons as the gas plasma region (in $t=5.5$ ns, this is the 
region $x>3$ mm), the gyro motion of the target ions that enter the 
gas plasma region is observed as the precursor in Fig.\ref{fig:spacetime}.
After the gyrating target ions 
reverse their direction ($t > 15.2$ ns 
\textcolor{black}{in Fig.\ref{fig:run2_1st}}), 
the accelerated gas ions gradually 
accumulate forward, leading to the steepening of density as well as magnetic field 
and the growth of the seed of a shock front ($t = 25.5$ ns), which 
\textcolor{black}{is} a 
seed-shock. The formation mechanism of a seed-shock in Runs 1 
and 4 is qualitatively the same as in Run 2. 

%

\subsection{Seed-shock formation: Run 3}

\begin{figure}[ht]
\includegraphics[clip, width=1.0\columnwidth]{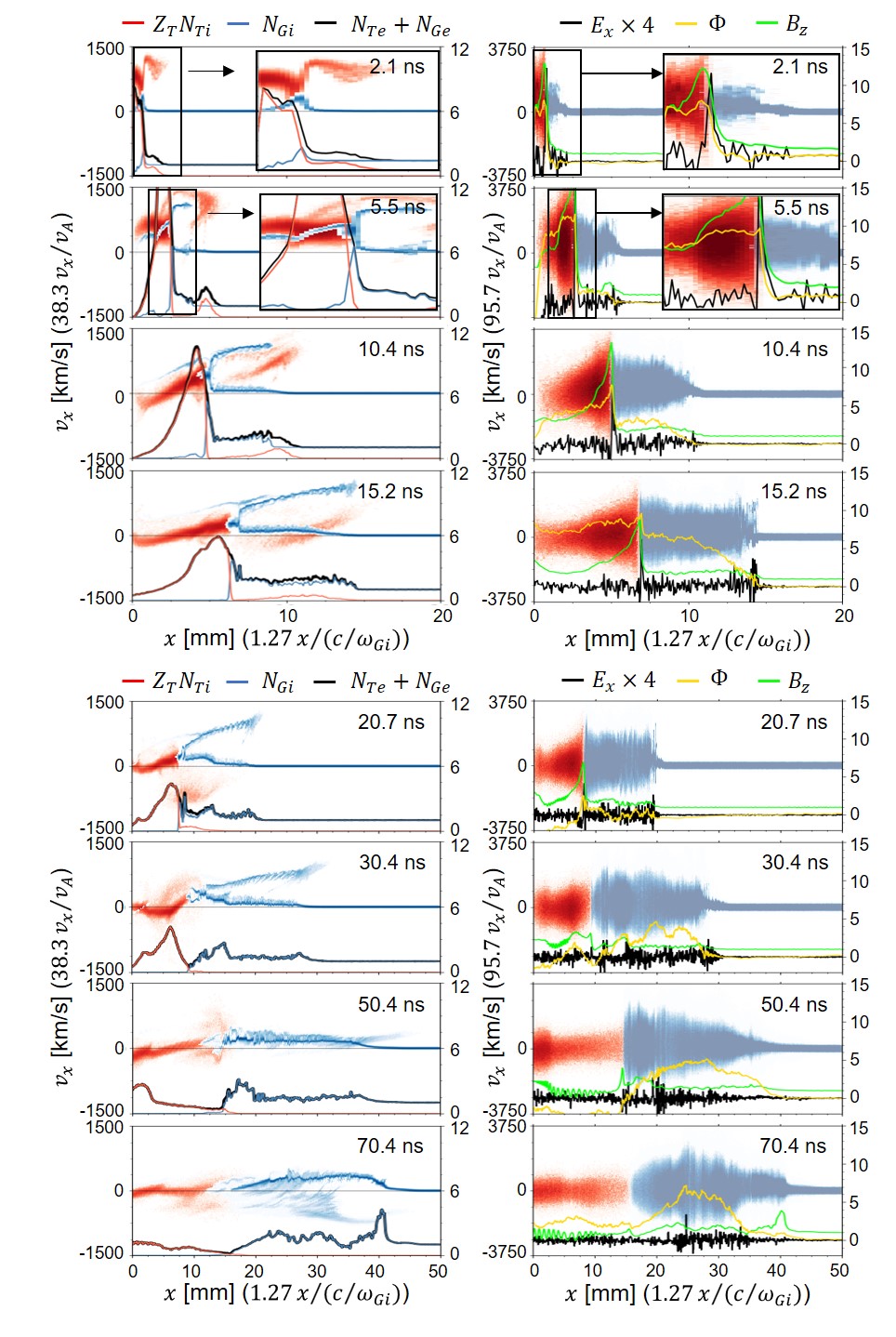}
\caption{\label{fig:run3_1st}The phase space distribution of target and gas 
plasmas, along with field profiles at various time points for Run 3. 
The format is consistent with that of Fig.\ref{fig:run2_1st}.}
\end{figure}
%
Fig.\ref{fig:run3_1st} depicts the same plot as Fig.\ref{fig:run2_1st} but for Run 3. 
While the initial behaviors of target plasma and gas plasma are similar to 
what are observed in Run 2 (Fig.\ref{fig:run2_1st}), more gas plasma is 
compressed and accelerated in the initial stage ($2.1-5.5$ ns). 
\textcolor{black}{
Since the target plasma is injected with the laser-generated magnetic field, 
both the flux of the injected target plasma and the magnetic flux it carries 
are proportional to the injection time. When the injection time is short, 
the target plasma disperses spatially due to velocity dispersion, which 
reduces the magnetic flux density carried by the target plasma, and the 
piston effect weakens rapidly (Run 2). On the other hand, if the injection 
time is sufficiently long, the target plasma with a constant magnetic flux 
density compresses the gas plasma for an extended period, resulting in a 
strong piston effect, and more gas ions are reflected by the magnetic 
piston (Run 3).
}
Therefore, more gas ions are reflected 
by the magnetic piston. This can be observed in $t \ge 5.5$ ns.
As a result, despite 
the accelerated target ions moving forward at a faster velocity than 
the reflected gas ions, their impact on the background gas ions has 
become relatively weaker compared to the case of Run 2. After the 
target ions move downstream through gyro-motion, the reflected gas 
ions form a precursor ($t \ge 20.7$ ns), accelerating background gas 
ions ($t=30.4-50.4$ ns) and creating the seed for a shock further ahead 
($t=70.4$ ns). The seed-shock formation mechanism in Runs 5 and 6 
is qualitatively the same as in Run 3.

\section{\label{sec:dependence}Parameter dependence of seed-shock properties}

\subsection{Dependence on $T_{in}$ and $B_0$}

\begin{figure}[ht]
\includegraphics[clip, width=1.0\columnwidth]{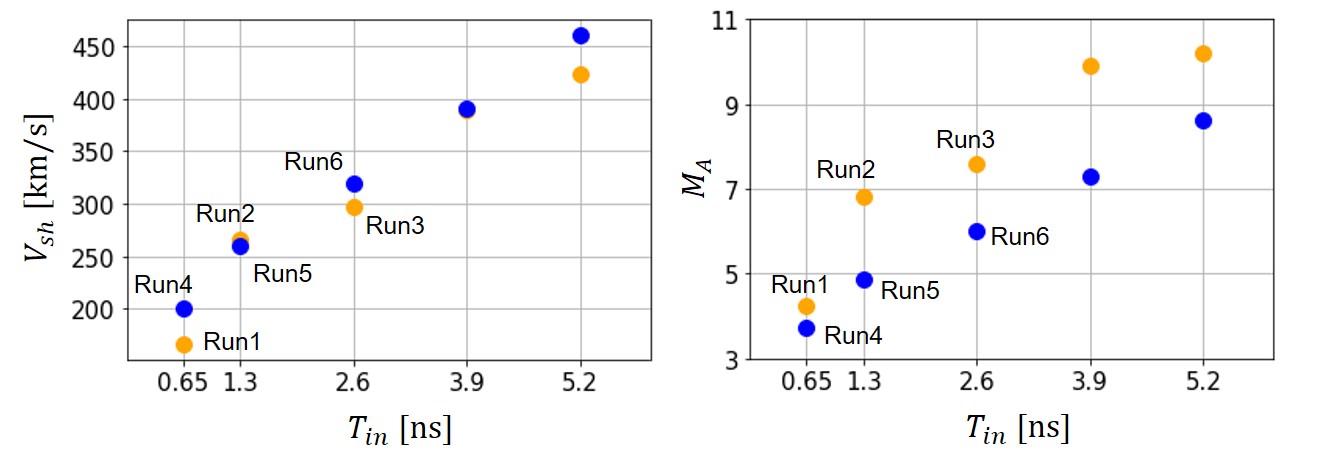}
\caption{\label{fig:ma}Velocity and Alfv\'{e}n Mach number of 
seed-shock for various $T_{in}$ and ambient magnetic 
field strength (yellow: $B_0=4.4$ T, blue: $B_0=6.0$ T).}
\end{figure}
%
The Alfv\'{e}n Mach number ($M_A$) and propagation velocity ($v_{sh}$) 
of generated seed-shock is plotted as a function of $T_{in}$ in 
Fig.\ref{fig:ma}. 
\textcolor{black}{
The velocity of a seed-shock is calculated as the peak-to-peak 
(from the first peak to the second peak) velocity 
of the electron density of a reforming seed-shock (Fig.\ref{fig:spacetime}).
}
In the figure, in addition to the Runs 1-6, the results 
for $T_{in}=3.9$ ns and 5.2 ns are also plotted. As $T_{in}$ increases, both $v_{sh}$ 
and $M_A$ of the seed-shock increase significantly. This is attributed to 
the suppression of target plasma deceleration by injecting target 
plasma for an extended duration, thereby maintaining a strong magnetic 
piston effect. It is noteworthy that 
a higher external magnetic field ($B_0$) leads to similar 
or a little higher $v_{sh}$, but $M_A$ decreases. 
However, in all runs, $M_A > 3$, indicating that the 
seed-shocks are supercritical \cite{tidman71}.

In Fig.\ref{fig:spacetime}, the high-density structure corresponding 
to the seed-shock exhibits periodic temporal variations. The periods 
of these variations are around 50-60 ns in Runs 1-3 and a little 
shorter than 50 ns in Runs 4-6. These values correspond to about 
one-quarter of the gyro period of gas ions in each run. Between the 
peaks of density maxima, there is a region where the density slightly 
increases, protruding forward. This region constitutes the foot region 
occupied by gas ions reflected by the seed-shock. These features are 
essentially the same as those of shock reformation 
\cite{biskamp72,lembege87,lembege92}, confirming that 
the seed-shock behaves similarly to the shock front of a supercritical 
shock.

\subsection{Other parameter dependence}

\begin{table}
    \centering

    \begin{tabular}{|c|c|c|c|c|c|c|c|}
    \hline
    Run \# & 3 & 3A & 3B & 3C & 3D & 3E & 6\\
    \hline
        $m_G/m_e$ & 100 & 50 & 200 & - & - & - & -\\
        $\Omega_{Ge}/\omega_{Ge}$ & 0.05 & - & - & - & - & - & 0.068 \\
        $\Omega_{Te}/\Omega_{Ge}$ & 6 & - & - & 3 & 12 & - & -\\
        $N_{Te}/N_{Ge}$ & 7 & - & - & - & - & 14 & -\\
    \hline
    \end{tabular}
    \caption{Values of parameters changed from Run 3 (first column).}
    \label{tab:my_label2}
\end{table}

\begin{figure}[ht]
\includegraphics[clip, width=1.0\columnwidth]{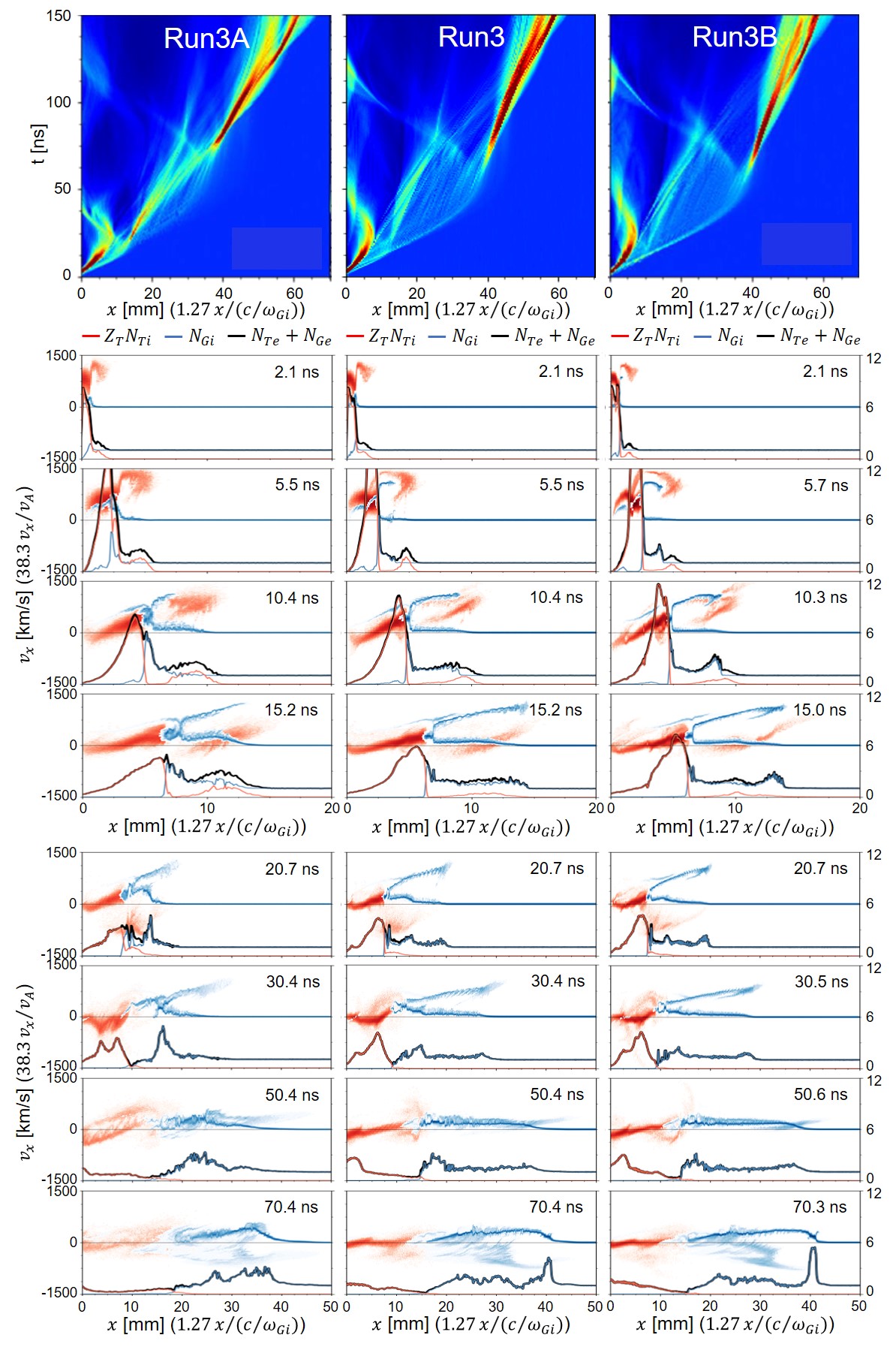}
\caption{\label{fig:rm}Mass ratio dependence of seed-shock 
formation. Run 3A: $m_G/m_e=50 ~(m_T/m_e=100)$, Run 3: 
$m_G/m_e=100 ~(m_T/m_e=200)$, and Run 3B: 
$m_G/m_e=200 ~(m_T/m_e=400)$, respectively.}
\end{figure}
%
Other parameter dependence is discussed here, 
with Run 3 as a reference. The parameters changed from Run 3 are 
summarized in Table \ref{tab:my_label2}. 
First, let us discuss the dependence on 
the ion to electron mass ratio. In Run 3, the mass ratio 
of the gas plasma was set to 100. Here, this value is changed 
to 50 (Run 3A) and 200 (Run 3B). However, the mass ratio of 
target ions to gas ions is kept fixed at 2 during this adjustment. 
Fig.\ref{fig:rm} shows the spatio-temporal evolution of electron
density and the time evolution of ion phase space and 
charge density profiles for each run. In all three runs, 
it is common for the precursor of reflected gas ions to eventually 
create the seed-shock. However, the relative density of target 
ions entering the gas plasma region in the early stages 
($t=5.5-15.2$ ns) becomes larger as the mass ratio decreases. 
This is due to normalizing particle velocities with the Alfv\'{e}n 
speed. As the mass ratio decreases and the Alfv\'{e}n speed 
increases, the thermal velocity of target ions also increases. 
As a result, ions with higher velocities enter the gas plasma 
region more frequently. In Runs 3 and 3B, the influence of target 
ions is suppressed, and the development of the system appears 
quite similar.

A similar behavior is observed when we change the ratio of electron 
cyclotron frequency to plasma frequency ($\Omega_{Ge} / \omega_{Ge}$) with other parameters being fixed. 
Increasing the frequency ratio results in a higher Alfv\'{e}n speed, 
and consequently, the thermal velocity of target ions increases, 
leading to more target ions entering the gas plasma region.

\begin{figure}[ht]
\includegraphics[clip, width=1.0\columnwidth]{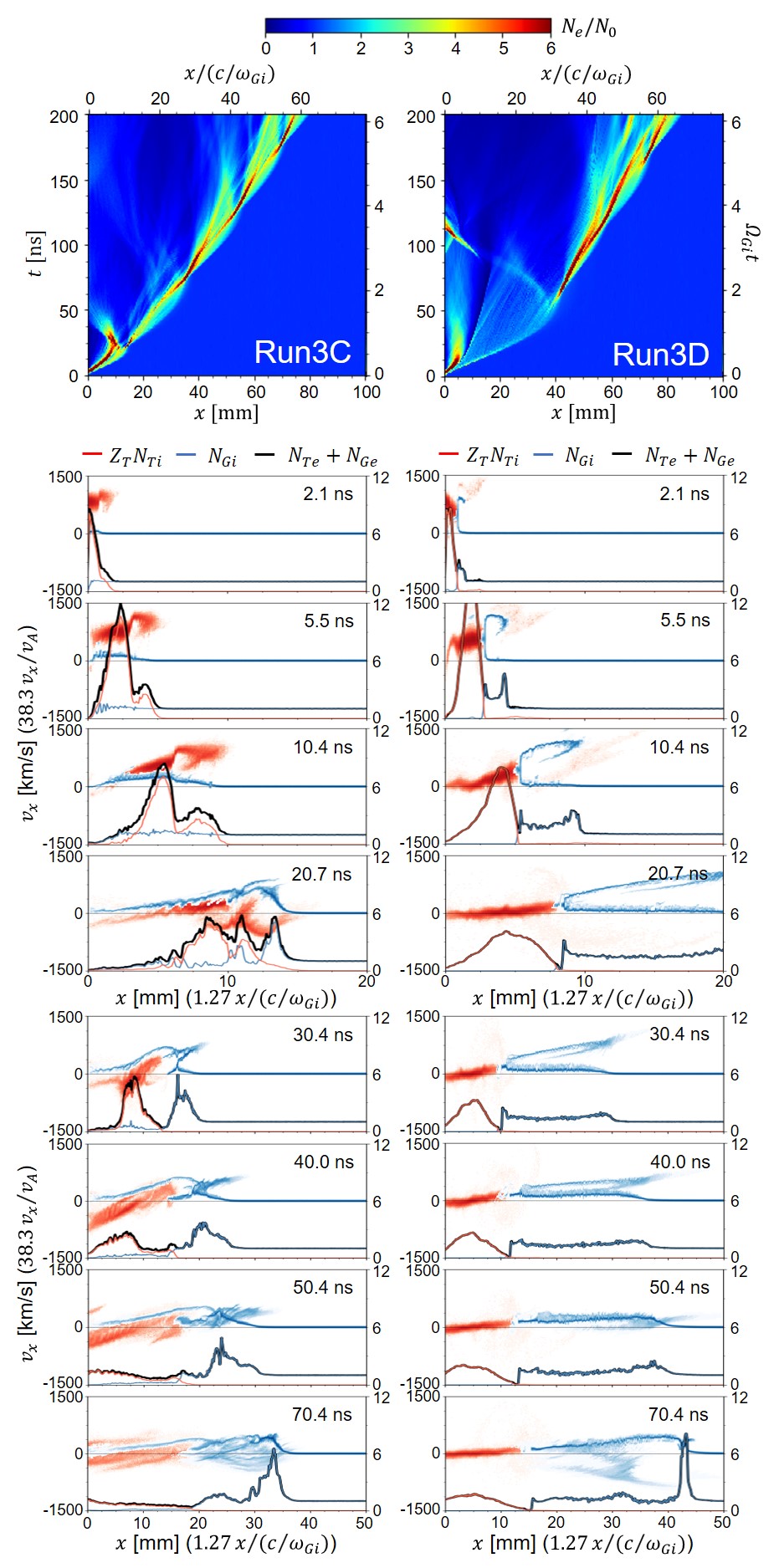}
\caption{\label{fig:mag}Dependence of seed-shock formation on 
the relative magnetic field strength of target plasma. Run 3C: 
$\Omega_{Te}/\Omega_{Ge}=3$ and Run 3D: 
$\Omega_{Te}/\Omega_{Ge}=12$, respectively.}
\end{figure}
%
The mechanism for seed-shock generation also varies depending on 
the relative strength of magnetic field carried by the target plasma. 
Fig.\ref{fig:mag} 
presents the results of calculations where the ratio of electron 
cyclotron frequencies between the target plasma and the gas plasma 
($\Omega_{Te}/\Omega_{Ge}$) is altered. In the case of low 
$\Omega_{Te}/\Omega_{Ge}$ (Run 3C: $\Omega_{Te}/\Omega_{Ge}=3$), 
more target ions enter the gas plasma to form the precursor. 
On the other hand, in the case of high $\Omega_{Te}/\Omega_{Ge}$ 
(Run 3D: $\Omega_{Te}/\Omega_{Ge}=12$), target ions do not 
penetrate the gas plasma region significantly. Instead, efficient 
reflection of gas ions contributes to the formation of the precursor. 
The propagation speed of seed-shock is faster in Run 3D.

\begin{figure}[ht]
\includegraphics[clip, width=1.0\columnwidth]{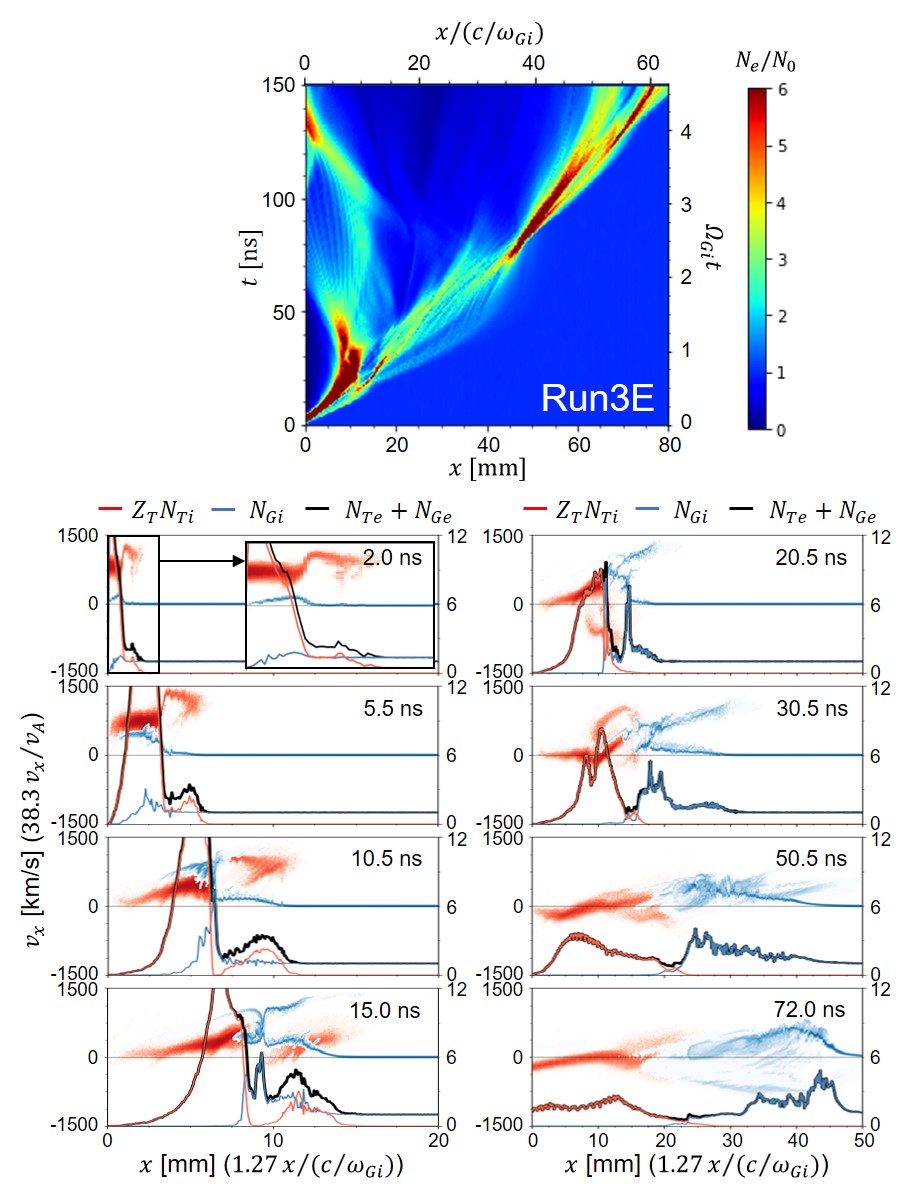}
\caption{\label{fig:den}Dependence of seed-shock formation on 
the relative electron density of target plasma for Run 3E 
($N_{Te}/N_{Ge}=14$).} 
\end{figure}
%
When we keep $\Omega_{Te}/\Omega_{Ge}$ constant and increasing the relative 
density of the target plasma (Run 3E: $N_{Te}/N_{Ge}=14$), it was observed 
that the piston effect by the target enhances and that the density 
of target ions entering the gas plasma region increases. 
Fig.\ref{fig:den} depicts the space-time plot of electron density 
and the temporal evolution of ion phase space and charge density 
profiles. In the space-time plot, the propagation speed of the 
seed-shock is higher compared to Run 3 (Fig.\ref{fig:spacetime}), 
indicating the strengthening of the piston 
effect from the target. Observing the temporal evolution of ion 
phase space and charge density profiles, the precursor by target 
ions has developed until $t=15.0$ ns due to the high density of 
target ions entering the gas plasma region. Simultaneously, at 
the boundary of the two plasmas, some gas ions are reflected, 
and the precursor originating from reflected gas ions grows, 
eventually surpassing the precursor from target ions 
($t=15.0-20.5$ ns). Ultimately, 
it grows into the seed-shock. Increasing the density ratio is 
expected to further enhance the precursor from target ions, 
contributing to the formation of the seed-shock.

\section{\label{sec:discussion}Discussions}

Up to this point, it has been established that the generation 
of the seed-shock requires a precursor, and there are precursors 
originating from target ions and reflected gas ions. It is 
understood that if a sufficient number of target ions enter 
the gas plasma region, they form the precursor, and for this, 
conditions such as a high relative density and low magnetization 
of the target plasma are suitable. Under the opposite conditions, 
reflected gas ions form the precursor. The actual experiment 
needs consideration of which condition is closer.

The ion-to-electron mass ratio and the ratio of electron cyclotron 
frequency to plasma frequency are both unrealistic in the 
simulations, leading to an overestimation of the Alfv\'{e}n speed, 
a constant for velocity normalization. Consequently, the thermal 
velocity of target ions increases, resulting in an excessive 
influx of target ions into the gas plasma region. Therefore, 
bringing these values closer to experimental values would likely
make it easier for the precursor from reflected gas ions to form.

On the other hand, the relative density and magnetization of the 
target plasma are expected to vary significantly in the 
three-dimensional spatiotemporal evolution of the system. The ion 
density of the target plasma immediately after laser irradiation 
is close to solid density, approximately $6 \times 10^{22} cm^{-3}$ 
for an aluminum target. The magnetic field generated on the 
target surface by the laser, known as the Biermann battery field, 
is around 100T \cite{sutcliffe22,umeda19}. The calculated 
magnetization of the target plasma 
from these values is on the order of 
$\Omega_{Te}/\omega_{Te} \sim 10^{-2}/\sqrt{Z_T}$. This is 
much smaller than the simulations here, implying that some 
target ions easily enter the gas plasma region. 
Assuming an initial velocity of target ions to be around 
1000 km/s, the gyro radius in a 100T magnetic field is on the 
order of 0.1-1 mm, depending on the valence of ions. This is 
roughly equivalent to the spot size of the irradiating laser. 
From these considerations, it is natural to assume that at 
least some target ions immediately after laser irradiation 
propagate through the gas plasma without sufficient magnetization.

In Ref.~\onlinecite{matsukiyo22}, a precursor was observed up to 
about 20mm from the target in an external magnetic field 
of 3.8T. If we attempt to explain this by considering the 
gyro radius of target (aluminum) ions, the initial velocity 
of target ions must be around 280$Z_T$ km/s. 
Yamazaki et al. (2022)\cite{yamazaki22} estimated the head speed of an 
aluminum target to be approximately 800 km/s, making it a 
reasonable value for $Z_T=3$. On the other hand, if we try 
to explain this using the gyro radius of reflected nitrogen 
gas ions, the initial velocity of reflected ions must be 
around 540$Z_G$ km/s. For example, the propagation velocity 
of the initial magnetic piston in Run 2 (up to 2.1 ns) is 
approximately 500 km/s. Assuming a similar velocity in the 
experiment, the speed of the gas ions reflected by the 
magnetic piston would be around 1,000 km/s, so 540 km/s 
($Z_G=1$) would be too 
small as the velocity of nitrogen ions when reflected by 
the initial magnetic piston. Values such as 1040 km/s ($Z_G=2$) 
or 1620 km/s ($Z_G=3$) would be more plausible. In other words, 
to identify the origin of the precursor observed in the 
experiment in Ref.~\onlinecite{matsukiyo22}, 
additional data and analysis are required.

We also propose another experimental design to clarify the 
origin of the precursor. Since the precursor reflects the 
gyro-motion of target ions or reflected gas ions, identifying 
the origin of the precursor should be possible by separating 
the spatiotemporal scales of both. Choosing light gas species 
with a small atomic mass is preferable for this purpose, and 
hydrogen gas is an ideal candidate. Hydrogen is known to ionize 
into singly charged ions, making it easier to estimate gyro 
radii and gyro periods. Yao et al. (2021)\cite{yao21} 
conducted experiments on 
magnetized plasma shock waves using hydrogen gas and CF2 
as the target, but they did not specifically focus on the 
initial stages of the system's temporal evolution. 
Helium, being capable of ionizing into singly or doubly 
charged ions, also provides an easily estimable scale for gyro-motion.

\textcolor{black}{
In this study, the effects of particle collisions have been 
neglected. The velocity of the ablation plasma (target ions) 
immediately after the shot in the Gekko XII experiment is 
estimated to be a little smaller than 1000 km/s. Using this 
value to estimate the Coulomb collision mean free path 
between a target aluminum ion entering the gas and the 
nitrogen gas ions, assuming a gas pressure of 5 Torr, 
the mean free path is larger than the size of the Gekko XII 
device. Similarly, the mean free path between the background 
ions and reflected ions in the gas is also larger than the 
size of the experimental region (a few cubic centimeters), 
indicating that ion-ion interactions can be treated as 
collisionless. On the other hand, the effect of 
electron collisions cannot be ignored. For example, referring 
to the calculation results here, if the electron temperature 
near the interface between the target and the gas plasmas 
is more than ten times ($\sim 100$ eV) that of the upstream gas 
electrons, the mean free path between the electrons and 
nitrogen ions in the gas is on the order of $\sim 0.1$ mm 
when the valence of ions is $Z_G=1$, and $\sim 0.01$ mm 
when the valence of ions is $Z_G=3$. This is comparable to 
the electron inertial length. However, the electron gyroradius 
is smaller, around $\sim 0.005$ mm (assuming a magnetic 
field of 5 T), so the gas electrons are magnetized (The Hall 
parameter \cite{lancia14} exceeds unity.). Regarding 
the target electrons, it has been reported that the target 
electrons expanding in vacuum are either unmagnetized or 
weakly magnetized, except near the target surface \cite{lancia14}. 
Therefore, the validity of the model used in this study, 
which assumes the injection of magnetized target plasma, 
requires further investigation. Note also that 
in \onlinecite{lancia14}, the laser spot size was 100 microns, 
which is an order of magnitude smaller than the value in the 
Gekko XII experiment, and the effects of the ambient gas 
were not considered, so careful verification is required to 
determine whether this applies to the current situation. 
Based on the above, the fundamental picture presented in this 
paper seems likely to occur in actual experiments as well. 
However, for quantitative discussions and phenomena where 
electron dynamics become significant, further investigation 
will be necessary, such as conducting calculations that 
explicitly incorporate collisional effects.}

Lastly, We would like to address the multidimensional effects. 
The 1D assumptions adopted in this study are quite strong. As 
mentioned earlier, in the experiment, the target plasma expands 
in three dimensions. In this study, we fixed the direction of 
the laser-generated magnetic field to the $Z-$direction, but 
in reality, the laser-generated magnetic field also has a 
three-dimensional structure. If we consider the boundary between 
the target plasma and the gas plasma as a multi-fluid plasma, 
various plasma instabilities could be excited. However, 
a 1D calculation may not accurately reproduce these instabilities. 
We recognize that multidimensional effects are a subject 
that should be carefully discussed in the future.

\begin{acknowledgments}
This research was supported 
by the JSPS KAKENHI Grant no 22H01287, 23K22558 (SM), 
22H01251, 23H01211, 23H04899, 23K22522 (RY), 20KK0064 (YK), and 22H00119 (YS). 
\end{acknowledgments}

\section*{Data Availability Statement}

The data that support the findings of this study are available from the corresponding author upon reasonable request.

\nocite{*}
\bibliography{aipsamp}

\end{document}